\documentclass[11pt,twoside]{article}
\usepackage{asp2014}

\aspSuppressVolSlug
\resetcounters

\begin{document}

\title{A GPU implementation of the harmonic sum algorithm}

\author{Karel~Ad\'{a}mek,$^1$ and Wesley~Armour$^1$}
\affil{$^1$Oxford e-Research Centre, Department of Engineering Science, University of Oxford, Oxford, OX1 3QG, UK;
\email{karel.adamek@oerc.ox.ac.uk}
}

\paperauthor{Karel~Ad\'{a}mek}{karel.adamek@oerc.ox.ac.uk}{0000-0003-2797-0595}{University of Oxford}{Oxford e-Research Centre}{Oxford}{}{OX1 3QG}{UK}
\paperauthor{Wesley~Armour}{wes.armour@oerc.ox.ac.uk}{0000-0003-1756-3064}{University of Oxford}{Oxford e-Research Centre}{Oxford}{}{OX1 3QG}{UK}

  
\begin{abstract}
Time-domain radio astronomy utilizes a harmonic sum algorithm as part of the Fourier domain periodicity search, this type of search is used to discover single pulsars. The harmonic sum algorithm is also used as part of the Fourier domain acceleration search which aims to discover pulsars that are locked in orbit around another pulsar or compact object. However porting the harmonic sum to many-core architectures like GPUs is not a straightforward task. The main problem that must be overcome is the very unfavourable memory access pattern, which gets worse as the dimensionality of the harmonic sum increases. We present a set of algorithms for calculating the harmonic sum that are more suited to many-core architectures such as GPUs. We present an evaluation of the sensitivity of these different approaches, and their performance. This work forms part of the AstroAccelerate project \citep{AstroAccelerateGit} which is a GPU accelerated software package for processing time-domain radio astronomy data.
\end{abstract}

\section{Introduction}
Detecting pulsars in time-domain radio astronomy using Fourier transform based techniques is a convenient and computationally efficient way to extract the faint periodic pulses from the noise in which they sit. However this technique, called periodicity searching, has some pitfalls. One of these is that the power contained in pulsar signal is spread into multiple harmonics in the calculated power spectra. The incoherent harmonic sum algorithm is one way to rectify this. 
In pulsar searches the observed time-series are first de-dispersed and then transformed using an FFT into frequency space. Then the harmonic sum is applied, which aims to sum the signal present and average out the noise. This is done by calculating partial sums of an increasing number of harmonics.

The harmonic sum algorithm sums the power that is spread across multiple harmonics back into a single Fourier bin. This increases the signal-to-noise ratio of detected pulsars and allows us to detect weaker pulsars as a result. 
The two-dimensional harmonic sum is also the next step after the Fourier domain acceleration search technique (FDAS) \citep{2002AJ....124.1788R}, which searches for accelerated pulsars. There is a GPU implementation of FDAS by \citet{2018arXiv180405335D}, there is also a GPU implementation of a two-dimensional harmonic sum for presto by \citet{PrestoGPU}.

The harmonic sum is given by equation
\begin{equation}
\label{eqa:prestoHRMS}
h(n)_\mathrm{H} = \frac{1}{\sqrt{H}}\sum_{i=1}^{H}x\left(\frac{ni}{H}\right)\,,
\end{equation}
where $H$ is the number of harmonics summed. The number of harmonics we need to sum is governed by the duty-cycle of the pulsar we are looking for.

The main problem with summing harmonics is that the peaks of the pulsar signal which we aim to add together can be, for higher harmonics shifted by number of frequency bins which are equal to the currently summed harmonic. The optimal harmonic sum performs all possible sums, which results in best signal-to-noise ratio. However, the shift in position of the pulsar signal for higher harmonics creates an unfavorable memory access pattern and also load balancing issues.

The harmonic sum algorithm is a standard in many software packages which process radio astronomy data, such as sigproc \citep{Sigproc} or presto \citep{Presto}. However none of these packages have the harmonic sum in a computationally accelerated form.

\section{Algorithms}
We have investigated a set of different harmonic sum algorithms, each algorithm has different properties and so can be used for different purposes. Some algorithms have good sensitivity, but suffer in performance and visa versa. We have implemented these harmonic sum algorithms on both the CPU and the GPU. The CPU versions of the harmonic sum are naively parallelized across the number of time-series processed. In the case of our GPU implementations the parallalization strategy depends on the harmonic sum algorithm under consideration.

We have compared our algorithms to the widely accepted harmonic sum based equation \ref{eqa:prestoHRMS}, which is widely used, for example in the SIGPROC software package. 

In this paper we introduce several harmonic sum algorithms which trade sensitivity for performance. The most obvious way to increase performance is to limit the number of sums examined by the algorithm. In the algorithm Max HRMS, this works by looking for a maximum at possible locations of the peak and adds this value to the partial sum for that harmonic sum.
\begin{equation}
\label{eqa:maxHRMS}
h(n)_\mathrm{H} = \frac{1}{\sqrt{H}}\sum_{i=1}^{H}\max_{1\leq j \leq i} \left(x\left(in+j\right)\right)\,.
\end{equation}

In freq. bin HRMS we only add values for bins which are integer multiples of the fundamental frequency, that is
\begin{equation}
\label{eqa:frbinHRMS}
h(n)_\mathrm{H} = \frac{1}{\sqrt{H}}\sum_{i=1}^{H}x\left(in\right)\,.
\end{equation}

Lastly in Greedy HRMS we recursively add the value in the appropriate bin or one of its neighbors depending what is bigger to the partial sum.
\begin{equation}
\label{eqa:greedyHRMS}
h(n)_\mathrm{H+1} = h(n)_\mathrm{H} + \max\left(x(Hn+j),x(Hn+j+1)\right)\\,,
\end{equation}
where j is increased by one if x(n+j+1) is selected.

\section{Results}
To measure the sensitivity of our algorithms we generated a time-series containing an artificial pulsar for which we have used the modified von Mises distribution. The same approach was used by \citet{2002AJ....124.1788R}. White noise was added using a pseudo random number generator with a normal distribution. The time-series contains 20 seconds of observing data with a sampling time $t_s=64\mu\mathrm{s}$.
We measure sensitivity using the signal-to-noise (SNR) ratio recovered by the algorithm for a pulsar of a given initial SNR in comparison to a standard algorithm based on equation \ref{eqa:prestoHRMS}. 

The pulsar's frequency could fall in between frequencies of the Fourier bins which result from the discrete Fourier transformation. This occurs when peaks of higher harmonics are non-integer multiples of the fundamental frequency. This results in a lower recovered SNR. The recovered SNR by different algorithms for such cases are shown in Figure \ref{fig:binSNR}. We see that the highest SNR detected is at the frequency  of the Fourier bins. There is the steep decline in recovered SNR for other pulsar periods. The frequency bin HRMS algorithm has for non-Fourier bin frequencies a sensitivity loss of more then 50\%. This is expected since the algorithm does not take into account the shift of the peak for higher harmonics. The Max HRMS algorithm has lower recovered SNR values, but they are consistent throughout the Fourier bin.

\articlefigure{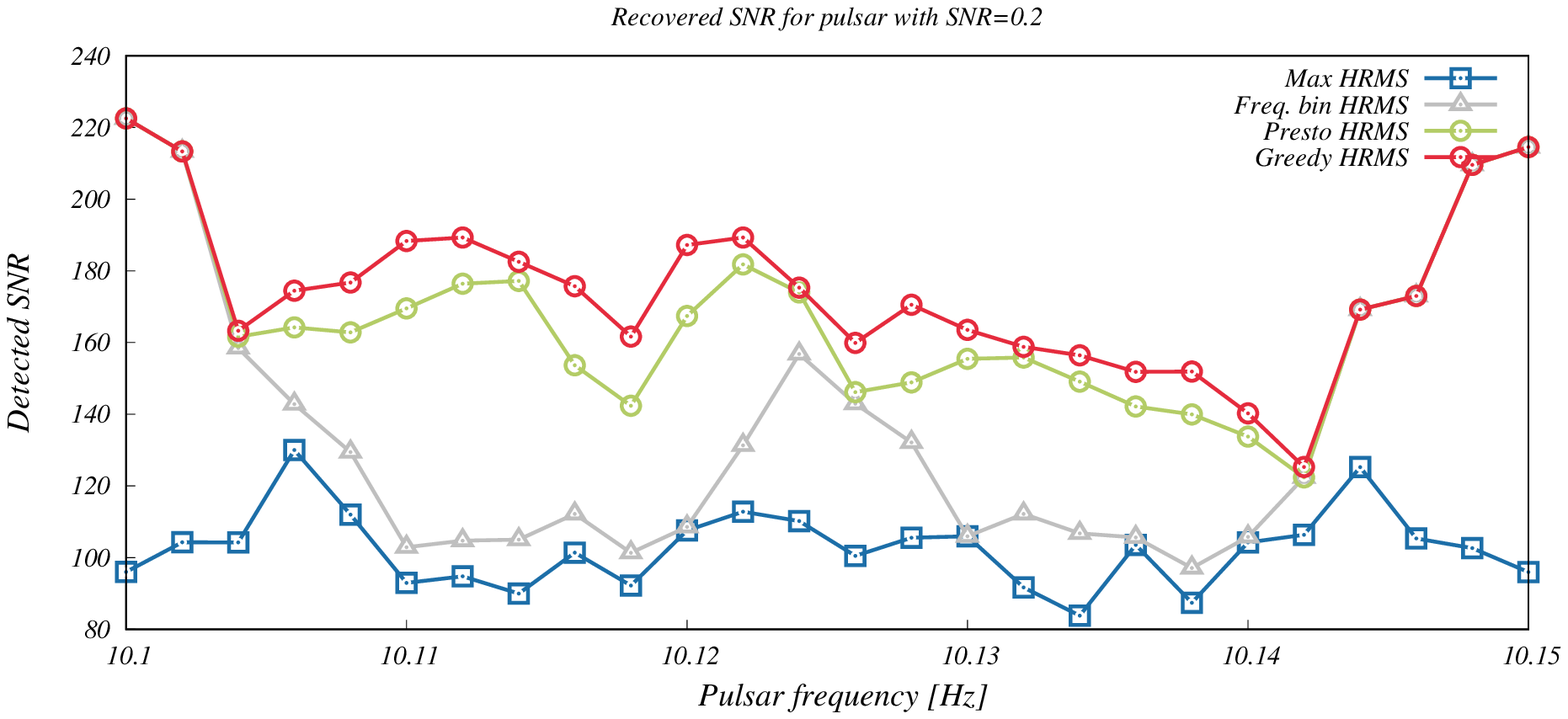}{fig:binSNR}{SNR recovered by different algorithms for pulsar frequencies which lie between discretised frequencies of the Fourier bins. Width of the bin is 0.05Hz.}

The averaged recovered SNR for a wide range of pulsar frequencies is shown in Figure \ref{fig:binaveragedSNR}. All presented algorithms recover consistent values of averaged SNR regardless of pulsar frequency. We see that standard HRMS and greedy HRMS has similar averaged SNR. The freq.bin HRMS algorithm has an averaged SNR loss of about 20\% and max HRMS algorithm of about 40\%. Lastly we present the performance of our implementations of the algorithms we have described here. For each algorithm we have implemented both CPU and GPU versions. The comparison in performance of the GPU versions to their respective CPU versions are shown in Table \ref{tab:Extimes}. This table also shows the speed-up of all GPU versions of these algorithms with respect to the harmonic sum algorithm based on equation \ref{eqa:prestoHRMS} which we have used as our testing standard. 

\articlefigure{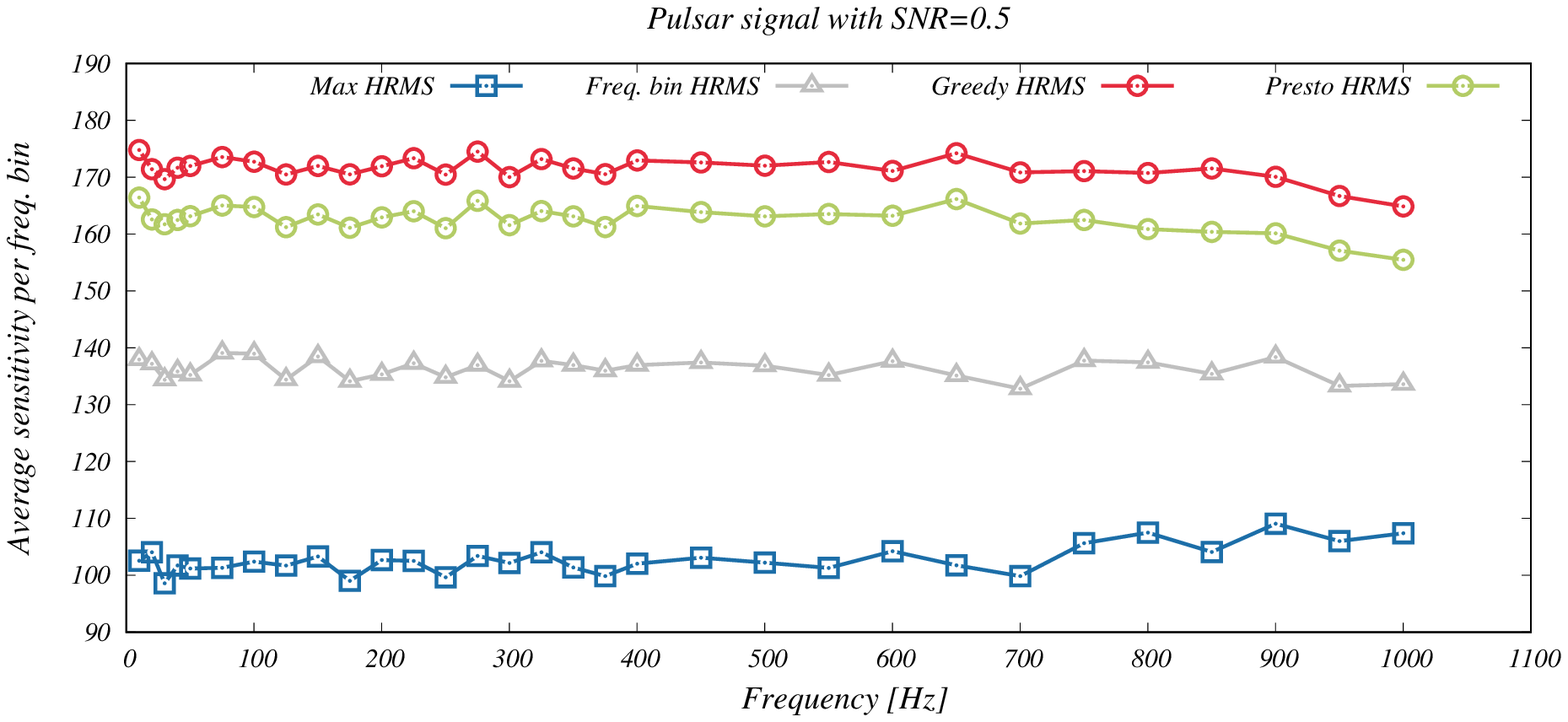}{fig:binaveragedSNR}{Averaged recovered SNR by different algorithms for a wide range of pulsar frequencies. All algorithms perform consistently on selected frequencies.}

\begin{table}[!ht]
\caption{Speed-up of GPU harmonic sum algorithms to their CPU counterparts and to our GPU implementation of the standard HRMS algorithm. }
\label{tab:Extimes} 
\smallskip
\begin{center}{
\small
\begin{tabular}{ccr}
\tableline
\noalign{\smallskip}
Algorithm & vs CPU & vs GPU Standard HRMS \\
\noalign{\smallskip}
\tableline
\noalign{\smallskip}
	Standard HRMS    & 250$\times$ & 1.0$\times$ \\
	Greedy HRMS   & 704$\times$ & 6.4$\times$ \\
	Freq. bin HRMS & 110$\times$ & 10.1$\times$ \\
	Max HRMS       & 177$\times$ & 3.9$\times$ \\
\noalign{\smallskip}
\tableline
\end{tabular}
}
\end{center}
\end{table}

\section{Conclusions}
We have presented the sensitivities of several algorithms for the calculation of the harmonic sum. We have implemented a set of these algorithms on both CPU and GPU hardware and compared the performance of the CPU vs. GPU implementation for each algorithm. We have also compared GPU implementations of other algorithms against the  performance of the standard HRMS algorithm. Our fastest algorithm is the freq. bin HRMS which has a sensitivity loss of approximately 20\% of the signal. A good ratio of sensitivity and performance is given by the greedy HRMS algorithm which has high sensitivity and also high performance, further work to explore this is underway. 


\end{document}